# Human-controllable AI: Meaningful Human Control

Chengke Liu

School of Foreign Languages, Anhui Agricultural University, Hefei, China

E-mail: ckliu@mail.ustc.edu.cn or ckliu@ahau.edu.cn

Wei Xu

Department of Psychology and Behavioral Sciences, Zhejiang University, Hangzhou, China weixu6@yahoo.com

**Abstract:** Developing human-controllable artificial intelligence (AI) and achieving "meaningful human control" (MHC) has become a vital principle to address these challenges, ensuring ethical alignment and effective governance in AI. MHC is also a critical focus in human-centered AI (HCAI) research and application. This chapter systematically examines MHC in AI, articulating its foundational principles and future trajectory. MHC is not simply the right to operate, but the unity of human understanding, intervention, and the traceablity of responsibility in AI decision-making, which requires technological design, AI governance, and humans to play a role together. The value orientation of MHC ensures AI autonomy serves human needs without constraining technological progress. The mode of human control needs to match the levels of technology, and human supervision should balance the trust and doubt of AI. For future AI systems, MHC mandates "human controllability as a prerequisite," requiring: (1) technical architectures with embedded mechanisms for human control; (2) human-AI interactions optimized for better access to human understanding; and (3) the evolution of AI systems harmonizing intelligence and human controllability. Governance must prioritize human-centered AI strategies: policies balancing innovation and risk mitigation, human-centered participatory frameworks transcending technical elite dominance, and global promotion of MHC as a universal governance paradigm to safeguard human-centered AI development. Looking ahead, there is a need to strengthen interdisciplinary research on the controllability of AI-driven autonomous systems, enhance the ethical and legal awareness among stakeholders, moving beyond simplistic technology design perspectives, and focus on the knowledge construction, complexity interpretation, and influencing factors surrounding human control. By fostering this transition in MHC, the development of human-controllable AI can be further advanced, delivering human-centered AI systems.

This chapter was funded by Scientific Research Project of Anhui Educational Committee (Humanities and Social Sciences) (2024AH040309) and Quality Engineering Project of Anhui Educational Committee(2023jyxm0213).

## 1 Introduction

In the era of AI, meaningful human control (MHC) has emerged as a key concept attracting significant attention. The concept of MHC was introduced by Filippo Santoni de Sio and Jeroen van den Hoven at Delft University of Technology. Primarily applied to Lethal Autonomous Weapons Systems (LAWS), MHC addresses the challenge of assigning responsibility when autonomous weapons make decisions (De Sio & Hoven, 2018). Its major principle is that humans must have the final say over the lethal choices made by these systems, thereby bearing the associated moral responsibility.

By definition, it does not simply refer to humans' operational authority over AI. Rather, in the critical stages of an AI system's operation, humans can participate in decision-making processes in an understandable, intervenable, and accountable manner. This ensures that the AI system aligns with human values, ethical principles, and social norms. Specifically, such control needs two things from humans. First, humans must understand how an AI system makes decisions. Second, when an AI system might cause big risks or drift away from its set goals, humans need to fix its behavior through effective intervention mechanisms that actually work.

In the development trajectory of AI, the necessity of MHC has become increasingly prominent, primarily reflected in two dimensions: human control over AI and accountability (León et al., 2021;Frenette,2023;Davidovic, 2023; Khomkhunsorn, 2025;Tsamados et al., 2025) . From a control perspective, with the advancement of AI's autonomous decision-making capabilities, especially in areas involving life safety and social equity, such as autonomous driving, medical diagnosis, and judicial assistance, ineffective human control may lead to AI systems making wrong decisions. This is usually caused by data bias, algorithm limitations, or unforeseeable scenarios, which can lead to irreversible consequences (Shneiderman, 2021). From the accountability dimension, AI behavior must ultimately be traced back to human agents, and MHC provides a clear basis for allocating responsibility (De Sio & Hoven, 2018; Calvert et al., 2020; Cavalcante et al., 2023; Nyholm, 2024). When there is a problem with the artificial intelligence system, if human beings can prove that they have fulfilled their reasonable control obligations at the critical stage, then there is a clear framework to define the accountability system. On the contrary, if human beings completely give up control, the ambiguity of responsibility will invalidate the accountability mechanism, thus undermining social trust.

MHC is closely intertwined with Human-Centered AI (HCAI), the key concepts of which provide an important perspective for understanding such control. HCAI emphasizes that the design of AI systems should focus on human needs, capabilities, and values, and its core concepts include interpretability, fairness, and the balance between autonomy and human dominance (Shneiderman, 2020; Xu et al., 2023; Schmager et al., 2023; Cape & Brereton, 2023). Among them, interpretability serves as the basis of MHC. Only when humans can understand an AI's decision-making process can they make effective interventions; Fairness requires that AI behavior does not violate human ethical norms and ensure that the goals of control is in line with the overall interests of society; The balance between autonomy and human dominance directly reflects the essence of MHC, that is, the independent decision-making of AI should always be under the reasonable supervision of human beings, which not only give full play to the efficiency advantage of AI, but also prevents human beings from losing control over technology.

Thus, MHC serves as a critical safeguard for the development of AI. It concerns not only the safety and reliability of technology but also the boundaries of human subjectivity and responsibility in the era of AI. Under the guidance of the design philosophy of Human-Centered AI, clarifying and implementing this concept will drive AI technologies to genuinely serve human well-being.

The MHC concept derived from lethal autonomous weapons has provided a useful reference for the effective integration of human-controllable AI and human control (De Sio & Hoven, 2018; Boardman & Butcher, 2019; De Sio et al., 2022; Flemischet al., 2023; Robbins, 2024).

The main goal of this chapter is to comprehensively and systematically analyze the core proposition of MHC, defining its connotations, developmental trajectory, core constituent elements, and realization pathways within the field of AI. It also thoroughly discusses the challenges, ethical and legal issues, and future directions, providing theoretical guidance and practical references for the development of AI. The scope of this chapter covers the evolution of human control of automation system from the historical perspective to the core elements of MHC in the field of AI; From the design idea to the related ethical and legal considerations; From the specific case study, to the challenges in the implementation process and the future development direction, the chapter strives to show the whole picture of MHC in all directions.

The HCAI approach emphasizes that AI should meet human needs, ensure human control, highlight human responsibility, and enhance human welfare. A key point in this approach is to ensure that AI systems are controllable by humans. In fact, there are different ways to achieve human control of AI, but MHC is one of the important practical paths. Therefore, this chapter will explore human-controllable AI from the perspective of MHC.

## 2 Historical Perspectives on Human Control of Automated Systems

Human control for automated systems has evolved alongside technological innovation since the Industrial Revolution (Hancock et al., 2013; Schwab, 2024). From early mechanical automation to today's AI systems, the human-AI interactions, control boundaries, and responsibility distribution have undergone profound changes. Tracing back to this historical track not only reveals the formation of MHC but also provides a reference for the AI governance.

### 2.1 From Automation to Autonomous Systems

Automation began with the Industrial Revolution, replacing human labor through mechanical structures like James Watt’s steam engine. The characteristic of this stage is “human-controlled automation”: machines, as extensions of human limbs, operate under direct and absolute command of humans through valves and switches (Lima & Belk, 2024). Systems like George Westinghouse’s pneumatic air brake accomplish complex, synchronized tasks, but their execution still heavily relies on human decision-making, with complete visibility and intervention capabilities.

The mid-20th century marked the rise of “programmed automation”. Electronics, represented by ENIAC, and Norbert Wiener's cybernetics introduced feedback loops using sensors and controllers. Industrial robots, such as the Unimate, executed pre-programmed sequences. Human control was layered: engineers designed offline logic, while operators monitored online. The system gained autonomous “execution” but retained transparent logic, allowing human override through code modification or physical intervention. Control remained largely human-centered.

In the 1980s, “adaptive automation” emerged. Microprocessors and sensors enabled early systems such as adaptive cruise control to make “local decisions” based on environmental data (such as radar). This introduced limited autonomy within predefined rules to adapt to dynamic conditions. However, humans remained the essential supervisors, and system autonomy was strictly limited, with logic still being understandable.

In the 21st century, “autonomous systems” driven by machine learning and big data have emerged. The “Stanley” vessel of Stanford University has achieved fully autonomous navigation, no longer relying on pre-programmed paths. Unlike its predecessors, these systems use deep learning to derive decisions from data, often resulting in an opaque “black box” effect (Rayhan, 2023). The victory of AlphaGo demonstrated AI decision-making that surpasses humans in complex environments, extending autonomy to critical fields such as healthcare. This evolution represents a fundamental shift from machines executing human instructions to systems making independent, data-driven decisions.

This progression poses a challenge to traditional human control. Autonomous systems operate faster than human reaction, and their decision-making is often difficult to understand, making direct supervision impossible. The evolution from automation to autonomous systems marks a profound shift in human-AI interaction (Sheridan, 2011). This transition challenges traditional control modes, as systems outpace human supervision and understanding. Therefore, this journey not only reflects technological advancement but also the redefinition of control, responsibility, and trust in an era of AI.

In a word, automation employs fixed rules and algorithms to perform predefined tasks, yielding deterministic outcomes, typically without AI. It shifts the role of humans from direct operators to supervisors, rather than replacing them. Autonomy, however, describes AI-driven systems capable of performing tasks independently without human intervention in specific operating contexts, while human controllability and authority must be retained as defined by the HCAI approach (Xu, 2020).

Consequently, the concept of MHC becomes essential. As autonomy increases, MHC ensures humans retain ultimate authority over ethical judgment and accountability (Robbins, 2024; Owolabi et al., 2024), defining the critical boundary between automation and truly autonomous systems.

## 2.2 Early Models of Human-Machine Interaction

The evolution of human-machine interaction models demonstrates a clear trajectory from automation to autonomous systems, while continuously redefining the boundaries of human control (Parasuraman, 1997). In mechanical automation, the interaction is characterized by a mapping between human actions and machine responses, such as Edison's phonograph, where control accuracy relies entirely on human physiology.

The electronic automation has driven the formation of the “program instruction mode”, the core of which is that humans preset the operational logic of machines through code or instruction sets, and machines strictly follow the rules to perform tasks The advancement of this mode lies in the fact that humans can control machines through abstract logical language, freeing themselves from the constraints of physical operations (Sheridan & Parasuraman, 2005); however, limitations also emerges. When environmental changes exceed the scope of preset rules, the system enters a rule blind zone and must rely on humans’ flexible judgment. This interactive logic provided an early prototype for the later concept of MHC.

Cybernetics promotes interaction into a “feedback control mode”, establishing bidirectional information flow between humans and machines (Mindell, 2002). Commercial airliner autopilot systems and Boeing’s fly-by-wire technology transformed humans from direct operators to parameter adjusters, focusing on setting boundaries rather than execution details (David, 2023). This model serves as a valuable reference for the control design of AI systems.

The AI era introduced “assisted decision-making models”, where machines evolved from executors to advisors, generating suggestions through data analysis, while humans retained ultimate authority. For instance, intelligent medical diagnosis systems and IBM Watson in cancer treatment embody this shift, where human control has shifted from operational details to value judgments (Fazelpour & Danks, 2021). However, when machine accuracy surpasses human experience, such a model will raise concerns about “over-reliance”.

Throughout the evolution from mechanical connection to decision-making assistance, every technological breakthrough has increased human trust in machines while reshaping control methods. The consistent adjustment of the “control boundary” indicates that as systems gain greater autonomy, the implementation of MHC has not only not diminished but has become even more crucial. This historical progression reveals that regardless of the degree of autonomy of the system, MHC always holds a central position in ensuring that technological development aligns with societal interests (Floridi et al., 2018).

## 3 Core Elements of Meaningful Human Control in AI

MHC is not an abstract concept, but rather a set of operational and implementable core elements. These elements involve the central role of humans in AI systems, the design of the technology itself, and the trust mechanisms in human-AI interactions. Only when these elements work together can it be ensured that AI systems always serve human values and avoid the risks of technological runaway.

### 3.1 Defining MHC

The concept of MHC originated within discussions about LAWS, where it was argued that such systems must remain under human direction and fully align with human intentions. Over time, MHC transcended its LAWS origins to become a fundamental principle in the ethical governance of artificial intelligence across diverse intelligent systems. MHC lacks a single definitive originator; instead, it rapidly evolved through debates surrounding the ethical, legal, and societal risks posed by autonomous weapons. The NGO ARTICLE 36 introduced the term in a 2013 report examining autonomy in UK weapon systems (Farrant & Ford, 2017). By 2014, MHC had become a central theme in United Nations discussions on LAWS under the Convention on Certain Conventional Weapons (CCW) (Perlinski, 2018; Riebe et al., 2020). Within the CCW framework, the Group of Governmental Experts (GGE) incorporated MHC, highlighting its necessity for AI ethics in multiple meetings (Riebe, 2023;Khajik, 2025). This adoption further advances and broadens the development of this concept, aiming to address potential legal and ethical challenges posed by autonomous weapons, such as those related to international law and humanitarian concerns.

Currently, MHC remains primarily an academic construct, not yet integrated into existing international humanitarian law, and lacks a universally accepted definition (Robbins, 2024). While conceptually appealing, its precise meaning and scope remain insufficiently clear, hindering its direct application as a policy or legislative instrument (Amoroso & Tamburrini, 2020). Stakeholders broadly agree on MHC as an ideal, but significant disagreements persist regarding its practical implementation. Different actors employ varied terminology to emphasize their specific priorities, resulting in no universal consensus on how to operationalize or apply MHC (Kwik, 2022).

However, MHC, as a unified concept, coordinates different viewpoints and represents a common position among supporters and critics of autonomous technology. Fundamentally, the international community agrees with the goals of MHC; The main challenge now is to determine the specific mechanism for achieving this goal.

Although there are different views on the specific meaning and implementation of MHC, people generally hold an optimistic attitude towards its potential. For example, the United States criticizes its subjectivity and ambiguity as operational barriers. Discussions surrounding MHC have significantly advanced its definition and practical application (Veluwenkamp, 2022; Davidovic, 2023; Abbink et al., 2024).

To establish MHC's philosophical viability, de Sio and van den Hoven drew upon John Martin Fischer and Mark Ravizza's "guidance control" theory (Fischer & Ravizza, 1998). This theory posits that moral responsibility requires rational control over the decision-making process leading to an action. They linked control directly to responsibility, identifying two essential conditions for MHC: first, the "reason-responsiveness" of the mechanism (the actor must act via a mechanism that recognizes and responds to reasons for or against the action, especially strong ones); second, the actor's ownership of the mechanism that triggers the action (Fischer & Ravizza, 1998). Achieving meaningful human control over AI systems is thus fundamental to assuming responsibility for their outcomes. These two conditions provide the theoretical underpinning for MHC and guide its practical realization.

Within the context of human oversight, MHC signifies the human capacity to make informed choices within adequate timeframes to influence AI systems, achieving desired results or mitigating harm now or in the future (Boardman & Butcher, 2019). Fundamentally, MHC contains two elements: "meaningfulness" and "control." "Meaningfulness" itself transcends technology; just as meaning in literature comes from interpretation, in technical systems it stems from the human intent shaping design and use, and the effective human command over the technology. Robbins (2024) deconstructs MHC by analyzing its three core words ("meaningful," "human," "control") to explore its multifaceted nature. "Meaningfulness" essentially denotes the essential prerequisites enabling humans to regulate machine operations and accept moral accountability. "Control" refers to the diverse methods humans employ to direct and manage AI systems (Christen et al., 2023). The term "meaningful" modifies "control" to stress its effectiveness: only control that effectively addresses the problem of suspended agency (subject suspension) and enables responsibility anchoring is truly "meaningful." If an AI system remains a tool under effective human control, instrumental nature will persist. The success of responsibility attribution hinges on whether humans achieve meaningful control over AI, which is intrinsically linked to the system's level of autonomy.

MHC functions more as a control philosophy than a specific, actionable theory. It defines the fundamental relationship between human controllers and controlled systems. This ensures that even without direct and constant operational control from human operators, moral responsibility and clear human accountability remain in place. However, when applied to concrete technologies like autonomous driving, MHC also aims to be translated into practical terms usable by engineers, designers, and policymakers. This translation involves defining the tasks, roles, responsibilities, and required capabilities of various human actors throughout the system's lifecycle (de Sio et al., 2022).

### 3.2 Human Agency and Decision-Making Power

Human agency serves as the fundamental principle of MHC, stressing that people must retain a leading position in shaping, implementing, and running AI systems, avoiding the role of passive technology users (Abbass, 2019). This agency operates in two key ways: first, the power to define goals, meaning the central aims of AI systems (like diagnostic precision in medicine or impartiality in court rulings) must be established by humans grounded in societal ethics and the common good (Ueda et al., 2024; Ejjami, 2024), not emerging independently through algorithms; second, final authority over crucial choices, especially concerning life safety and social justice, where people must keep the ability to reject or alter AI outputs (Christen et al., 2023).

Consider autonomous vehicles: their ethical priorities (for instance, whether to prioritize passenger or pedestrian safety during a crash) must be set by humans via societal agreement, not unilaterally chosen by programmers. Individuals from diverse cultures hold different preferences regarding ethical dilemmas in self-driving cars. An AI system's value framework must stem from collective human societal choices, not the technology's inherent logic (Gabrie, 2020).

The distribution of decision-making authority represents a practical expression of human agency (Wagner, 2019). Within AI systems, the allocation of decision-making power should follow the hierarchical authorization method. For low-risk, routine activities (such as data input and basic customer support), AI can be granted full decision-making autonomy; for moderate-risk tasks (like e-commerce suggestions and content moderation), AI may propose decisions, but human supervision is mandatory; for high-risk operations (including controlling surgical robots and deploying autonomous weapons), AI must function solely as a support tool, with humans making the ultimate determination.

The essence of this hierarchical authorization is to guarantee the effectiveness of human authority，which means that human choices must truly shape the ultimate outcome of AI systems. (Crootof et al., 2023). For example, in an AI-driven diagnostic tool, if a physician disputes the AI's conclusion, the system must offer a mechanism for modification, and the revised result must be implemented; conversely, if the system's design blocks human decisions from taking effect (e.g., algorithms overruling human adjustments), it undermines the core of MHC.

In addition, protecting human decision-making relies on the principle of ability matching. The interface and operation process of AI systems must match human cognitive patterns to prevent the weakening of human decision-making ability due to overly complex designs or excessive information demands (Véronneau & Cimon, 2007).

### 3.3 Transparency, Explainability, and Accountability in AI Systems

The technical foundation of MHC interconnection is transparency, interpretability, and accountability, which together constitute its core elements.

Transparency means humans can see the design logic, data sources, and operational processes of AI systems (Felzmann et al., 2020; Andrada et al., 2023). This requires developers to reveal core algorithm principles , users to know when they are interacting with AI instead of people, and regulators to have access to system operation logs.

A lack of transparency directly creates a control vacuum (Abrams et al., 2019). For instance, Amazon halted its AI recruitment tool in 2018 because it implicitly discriminated against female candidates. This occurred because the system's training data and algorithmic feature extraction logic weren't adequately disclosed, which stops humans from spotting the bias quickly. This situation proves that without transparency, humans cannot control AI.

Explainability is a further requirement built on transparency, focusing on whether the logical chain of AI decisions can be understood by humans (Balasubramaniam et al., 2023). For simple rule-based AI ,

decisions can be directly explained through ‘if-then’ logic; however, for complex deep learning models , the difficulty of explanation significantly increases. To tackle this, scholars proposed local explainability, meaning it’s sufficient to explain the basis for a single decision rather than understanding the entire model (Ribeiro et al. 2016;Lundberg et al., 2020; Hassija et al., 2024).

Explainability is necessary for humans to intervene in AI decisions (Rajendra & Thuraisingam, 2025; Amann et al., 2022). In the legal system, if an AI sentencing tool suggests a heavy sentence for a certain case, the judge needs to know if it’s based on “crime circumstances”, “prior convictions”, or other factors, otherwise, it is impossible to judge the validity of the recommendation.

Accountability represents the ultimate aim of transparency and explainability, defining who bears responsibility for AI system behavior (Williams et al., 2022; Novelli et al., 2024). Responsibility should extend to all involved parties: developers (for design flaws), deployers (for inappropriate applications), and users (for operational mistakes). For example, if an autonomous vehicle accident occurs due to an algorithm flaw, the developer is liable; if it results from the user failing to take control promptly, the user is liable.

Traceability is central to accountability (Kroll, 2021). AI systems must maintain detailed decision logs that document input data, algorithm versions, and any human interventions for each decision. This ensures precise responsibility assignment when problems emerge.

3.4 The practical framework of MHC: tracking and tracing

Fischer and Ravizza’s guidance control theory outlines the conditions for human moral responsibility, focusing on the link between decision-making mechanisms and the subject (Fischer & Ravizza, 1998). Classical control theory emphasizes causal links between controller and controlled system, but MHC prioritizes abstract coordination—where system behavior aligns with the controller’s moral reasons, intentions, and goals (de Sio et al., 2022).

A core condition is the “reason-response” requirement: the decision-making mechanism must respond to moral inputs, adapting system behavior to environmental moral features (e.g., human mental states or external world characteristics) (de Sio & Hoven, 2018). This establishes a moral motivation link between system behavior (e.g., human operators or decision-support interfaces) and the subject, particularly for ethical motives that could cause harm.

De Sio and Van den Hoven (2018) expanded this with two conditions for intelligent systems under human control: “tracking” (system recognizes and executes human moral intentions) and “tracing” (anchors human responsibility to the control chain). Tracking is MHC’s first necessity: systems must capture human moral reasons, even when human influence on the world is indirect (e.g., via systems, creating distance between action and outcome). This indirect control often takes implicit or chain-like forms, making accurate tracking critical.

For meaningful control, the decision-making mechanism must accurately capture human moral reasons across contexts and ensure results are verifiable. This “tracking” is foundational for bridging the responsibility gap, as it connects system behavior to human motives.

Effective tracking requires systems with robust situational recognition, enabling them to accurately perceive environmental conditions, identify relevant features, and extract qualifying targets. While De Sio and Van den Hoven did not specify tracking subject characteristics, they established a key constraint: subjects must be humans possessing free will. This implies that the core purpose of tracking is to identify traces of human behavior or control within complex system chains, crucially uncovering the human will, such as that of designers, legislators, or decision-makers, behind the system. Thus, tracking may not manifest as direct human action; even system behavior must stem from human intervention. Furthermore, MHC does not guarantee value alignment, as systems could still adopt harmful orientations, necessitating vigilance. MHC is therefore necessary but insufficient.

However, tracking is not value neutral. De Sio and Van den Hoven (2018) argue that it requires systems to respond to human moral reasons, imposing a normative demand: design must reflect relevant principles, norms, and values. Consequently, even if tracking is deemed essential for MHC, disagreements persist about its achievement in specific cases.

Building on Fischer and Ravizza’s “ownership condition” (requiring agents to understand and accept their moral decision mechanisms), De Sio and Van den Hoven propose traceability as MHC’s second necessary condition. This means any system action under MHC must be traceable to relevant human subjects and their moral judgments during design or operation. Specifically, at least one human involved in the system’s lifecycle must: (1) understand or be capable of understanding the system's functions and potential impacts, and (2) understand or be capable of understanding the reasonable moral responses others might have to those impacts and their role in them. If system behavior cannot be traced to such human comprehension and acceptance, it lacks MHC, regardless of intelligence.

Traceability allows holding actors responsible for outcomes if their prior actions relate to later responsibility determinations, even if they weren't culpable at the time of action (e.g., someone encouraging drinking might share liability in a drunk driving case). De Sio and Van den Hoven (2018) illustrate this with scenarios like Jim's execution dilemma and pilots using Collision Avoidance Systems.

Applying traceability to autonomous systems presents challenges: it must extend to situations involving (1) multiple human subjects and (2) non-intelligent subsystems contributing to outcomes. Responsibility allocation between operators and designers is debated; if operators aren't morally responsible or at fault, accountability may shift to upstream actors or designers who failed to properly understand the system (de Sio & Hoven, 2018).

Therefore, the second criterion for MHC over autonomous systems is established by examining the traceability between the system's decisions and the technical and ethical understanding of personnel involved in its design and deployment.

### 3.5 Trust and Human Oversight in AI Decision-Making

Trust and human oversight are the core elements of MHC at the practical level, forming a dynamic balance: trust is the foundation of human-machine collaboration, while oversight is the guarantee against trust abuse (Blanco, 2025). Trust in AI decision-making is a form of rational trust that is based on empirical evidence of system performance, reliability, and ethical consistency, rather than blind reliance (Lee & See, 2004). The formation of this trust requires three conditions: first, the decision accuracy of the AI system needs to be verified in long-term practice ; second, the behavior of the system is stable ; third, the system can promptly alert of its own limitations .

Rational trust can enhance the efficiency of human-AI collaboration, as seen in financial risk control, where staff focus on AI-flagged high-risk transactions (Hoc, 2000). But trust needs to be maintained within a reasonable range: excessive trust can lead humans to give up supervision, while insufficient trust can weaken the value of AI.

Human oversight is a key balancing mechanism, classified by risk level: “real-time oversight” for high-risk scenarios , “regular review” for medium-risk applications , and “post-hoc traceability” for low-risk systems. Effective oversight requires reducing automation bias, where humans neglect their judgment due to over-reliance on AI (Romeo & Conti, 2025), and ensuring supervisor capability through adequate training on AI principles and limitations (Tsamados et al., 2024).

This dynamic balance represents continuous verification of AI controllability (Kieseberg et al., 2023). Trust increases when AI performs as expected, which allows reduced oversight intensity; abnormal situations or new contexts require reduced trust and strengthened oversight. This flexibility is essential for MHC in a rapidly evolving technological landscape. Trust and oversight are interdependent elements, which ensures human agency and effective control throughout the AI lifecycle.

## 4 Designing for Meaningful Human Control

MHC is not a natural result of technological development, but rather requires active design intervention to achieve. From the interaction of the user interfaces to the closed-loop design of the system, and then to the implementation of explainable technologies, every stage of design directly affects human control over AI systems (Shneiderman, 2020;Cavalcante et al., 2023; Xu, 2024). Only by embedding the concept of "control" into the full lifecycle design of AI systems can it be ensured that technology remains under reasonable human control.

### 4.1 Designing AI Interfaces for User Control and Feedback

The user interfaces of AI systems act as the primary channel for human-AI interaction, and their design quality fundamentally influences how easily and effectively users can exert control (Rzepka & Berger, 2018;Abedin et al., 2022;Xu, 2024). A well-designed AI interface should prioritize user control, which will balance efficiency with the ability for users to clearly understand the limits of their control, conveniently exercise their control options, and receive prompt system feedback (Shneiderman, 2020).

The core of user interface design lies in creating layers of control granularity. Users have different control needs depending on their role (Cavalcante Siebert et al., 2023): everyday users might only require basic commands like “start/stop” and “accept/reject”; expert users need advanced functions such as parameter adjustment and rule modification; whereas regulators may require administrative controls like accessing system logs and auditing decision histories. For instance, a user interface of an intelligent medical diagnosis system should offer patients basic controls such as “view diagnosis conclusions” and “request human review”, provide doctors with advanced controls like “adjust model weights” and “add clinical parameters”, and grant hospital administrators oversight controls including “view system accuracy” and “review abnormal cases”. This layered approach prevents overwhelming ordinary users with unnecessary complexity while guaranteeing that professional users possess the necessary depth of control (Sartori & Theodorou, 2022; Jubair et al., 2025).

It is equally vital that the immediacy and clarity of feedback mechanisms are ensured. The interface must promptly communicate any AI decision or system state change to users, and the feedback presented should be intuitive and align with human cognitive patterns (Bader & Kaiser, 2019). For example, when an intelligent recommendation system alters its strategy, it should explicitly inform users that “this recommendation is based on your browsing history and the preferences of similar users”. If an autonomous driving system switches to manual mode due to sensor failure, it must alert the driver through multiple sensory channels (visual and auditory) and display the prompt “please take control of the vehicle immediately”.

Additionally, the interface design should have simple emergency controls. In high-risk situations, users require the ability to quickly regain control during critical events.(Van de Walle & Turoff, 2008). For example, surgical robot interfaces should feature a highly visible “emergency pause” button, which can ensure its activation overrides any autonomous system actions. Smart factory monitoring interfaces should enable administrators to instantly halt AI control of production equipment with a single click, preventing accidents caused by system errors. This design maintains AI functionality while providing an essential safety measure for unexpected situations.

### 4.2 Control Mechanisms in Human-in-the-Loop (HITL) Systems

Human-in-the-Loop (HITL) (Feng et al., 2016;Muhammad, 2021 ) systems integrate humans into the AI decision-making process, forming a collaborative control chain. Its core control mechanisms encompass dynamic authorization, collaborative decision-making, and feedback learning, which interact to form a stable human-AI collaborative relationship.

The dynamic authorization mechanism automatically adjusts the allocation of human-AI control authority based on system status and environmental changes. (South et al., 2025;Tsamados et al., 2025). AI autonomously processes routine low-risk tasks, such as data preprocessing. But once abnormal situations are detected (such as missing data or low confidence), control will be seamlessly transferred to humans, while providing relevant data and recommendations. For example, the smart grid dispatch system operates autonomously when the load is stable, and immediately issues an alarm to the manual dispatcher and grants operational authority when a fault occurs. This mechanism ensures that key nodes are controlled by humans, leveraging the efficiency advantages of artificial intelligence while retaining ultimate decision-making power. This is the core cornerstone of achieving effective control.

The collaborative decision-making mechanism fully leverages the advantages of both humans and AI: AI excels in data processing and computation, while humans provide ethical judgment, common sense cognition, and crisis management capabilities (Etuk & Omankwu, 2025). Effective control requires a clear division of labor: AI provides data-driven solutions and analysis, while humans retain the final decision-making power and value-based judgments. For example, in the judicial system, AI can provide sentencing recommendations based on precedents, but judges need to consider subjective factors such as malicious intent and social influence to make the final decision. This division of labor not only avoids being completely dominated by AI algorithms but also reduces the cognitive load on humans, which can ensure that human judgment always occupies the core position of control.

Feedback learning utilizes human-AI interaction to optimize AI systems. The HITL system records human corrections (acceptance, modification, rejection) to the output results of AI and uses this data for model optimization (Joshi, 2025). For example, by analyzing how customer service agents adjust AI responses, weaknesses in emotional understanding can be identified, thereby improving natural language processing models in a targeted manner. This continuous cycle, based on human feedback, model optimization, and enhanced control, enables AI to adapt to human control modes, gradually improving system efficiency and human oversight quality over time.

Together， these mechanisms collectively construct a framework in which human control is not merely present but is substantive, adaptive, and effectively integrated, which are crucial for designing systems where human agency is genuinely meaningful.

### 4.3 Designing Explainability into AI for Enhanced Control

Explainability is not a feature that can be added afterwards, but must be integrated into the technical architecture of AI systems from the design stage as a fundamental ability to support human control (Ahmed et al., 2022). This “built-in explainability” design uses technical means to reduce the “black box” nature of AI decisions, enabling humans to exercise effective control based on understanding.

To ensure that AI systems are not only powerful but also controllable, designing explainability into their core architecture is essential (Dwivedi et al.,2023;Chamola et al., 2023). The explainability of the model is a fundamental requirement(Ahmed et al., 2022). When performance requirements are met, models with high explainability should be prioritized. These simple models are easier to understand and debug because their decision-making process is transparent and intuitive. For example, decision trees can clearly display the sequence of conditions that lead to decisions, making it easier for users to trust and verify the system. Although complex models may have higher accuracy in certain scenarios, their lack of explainability poses significant challenges to oversight and control, especially in high-risk applications.

Another key aspect is “user customized explanation design”. Explanations must be carefully customized based on the specific needs and professional knowledge of users (Giovannoni, 2024). For example, ordinary users may need a concise and easy to understand summary when interacting with medical AI. In contrast, medical professionals may require detailed information such as model confidence, input features, and alternative diagnostic options. Customizing explanations to fit users’ backgrounds makes sure they are easy to understand and use. This builds trust and helps people make decisions with enough information.

The core of effective control lies in real-time explanation and intervention(Wu et al., 2023; Gyory et al., 2022; Rajendra & Thuraisingam; 2025). Explanation should not remain at the level of “post hoc explanation”, but should be directly related to control operations. When AI systems make decisions, clear real-time reasoning explanations should be provided synchronously. For example, when an autonomous vehicle detects a pedestrian, it not only needs to brake, but also needs to explain the reason. This enables human operators to understand the situation, intervene when necessary, and provide feedback to optimize the system. This continuous cycle constitutes a closed-loop control system that enables humans to dynamically guide AI decisions, correct errors, and ensure consistency with human intentions. This method enhances human oversight and control, making AI systems more reliable and trustworthy in critical applications.

4.4 Effective human control

The practical implementation of MHC is a subject of significant scholarly inquiry across fields, notably by the European interdisciplinary team led by Boardman and Butcher (2019), alongside others (Cavalcante et al., 2023; Hille et al., 2023; Boutin & Woodcock, 2024). Their research highlights that while MHC is fundamentally rooted in moral ethics and legal perspectives, its realization demands the incorporation of “Effective Human Control” (EHC) within system design. The vision of MHC can only be fully achieved through the mutually reinforcing integration of MHC (focusing on strategic guidance) and EHC (focusing on operational execution), both of which are vital components of human control in autonomous AI.

According to Boardman and Butcher (2019), MHC encompasses two primary elements:

Legal: Mandates adherence to International Humanitarian Law and upholds accountability for military operations in both physical and information domains.

Moral/Ethical: Provides the essential ethical foundation for military decision-making and actions across physical and information domains.

The essence of MHC, whether through binding legal requirements or guiding moral principles, is the pursuit of “technology for good.” This ensures technology’s evolution and application consistently follow a beneficial trajectory, consciously adhering to ethical, legal, and social constraints, and reflecting the fundamental principle that humanity is the ultimate end. This core purpose enables MHC to gain widespread acceptance and motivates the search for concrete implementation strategies.

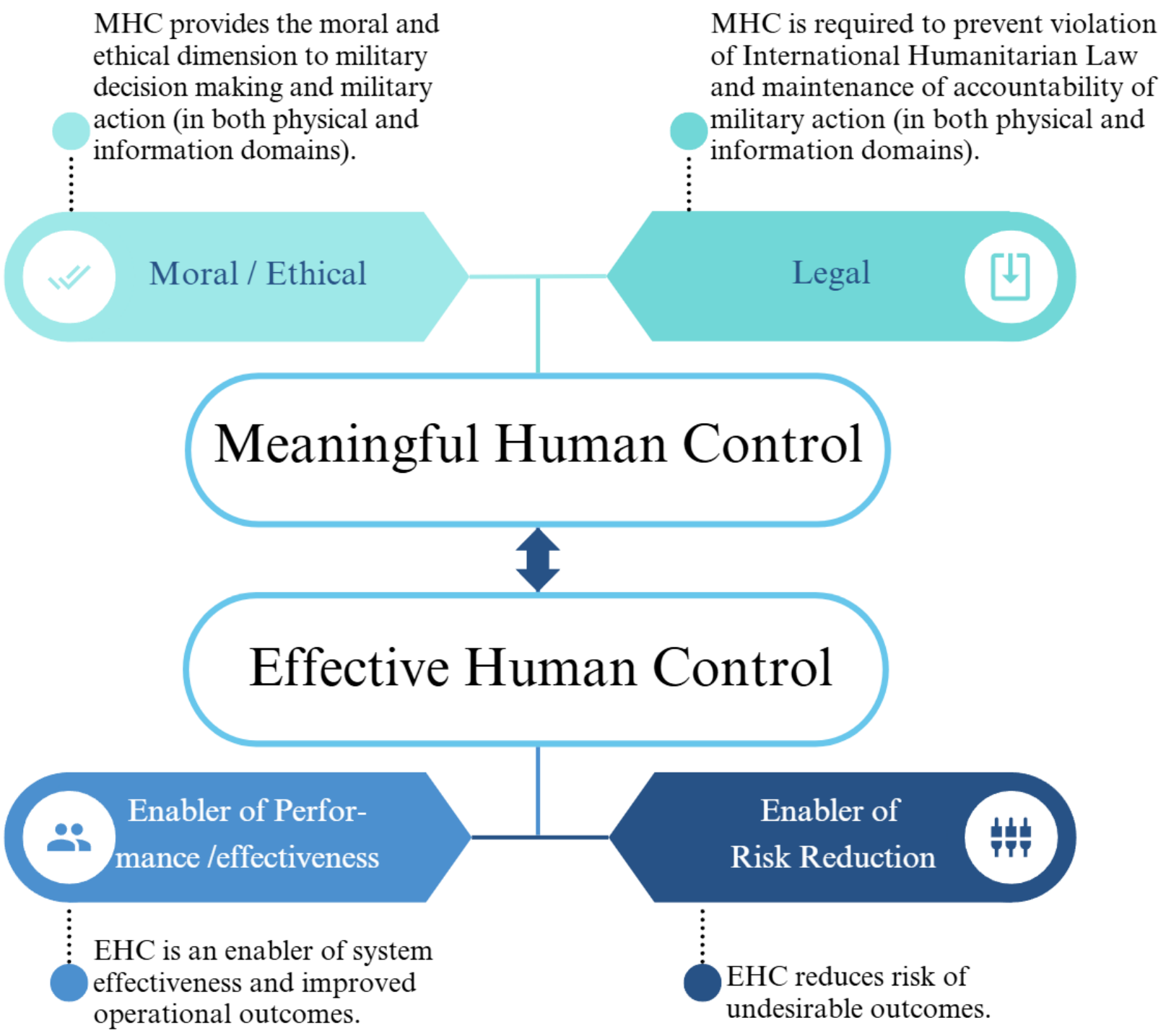

Figure 1 Meaningful Human Control and Effective Human Control (Boardman & Butcher, 2019)

The necessity of EHC stems from the fact that humans and machines working together towards a shared goal are more effective than either operating independently. According to Boardman & Butcher (2019), EHC, which specifically involves human involvement in decision-making, acts as a key enabler in two ways: it enhances operational performance and effectiveness, and it reduces risks and undesirable outcomes, thereby strengthening system resilience.

Figure 1 reveals that MHC begins from a macro perspective, highlighting the necessity and significance of human control and placing AI autonomous systems within ethical and legal frameworks. EHC, however, works from a micro perspective, concentrating on the technical realization of MHC's concepts to minimize risks and boost system efficiency. Considering both aspects together is crucial. Concentrating only on efficiency and system performance might unintentionally erode meaningful control. Conversely, requiring excessive user input for certain decisions can result in delays, biases, or mistakes, negatively impacting business efficiency (Boardman & Butcher, 2019).

## 5 Ethical and Legal Considerations for MHC

With the application of AI in highly autonomous technological fields, MHC is no longer just a simple technical issue, but also involves complex ethical judgments and legal regulations (Mecacci et al., 2024). As AI systems gain more and more decision-making autonomy in fields such as healthcare, transportation, and justice, how humans can maintain the benefits of technology while maintaining ethical boundaries and clarifying legal responsibilities has become a core issue that needs to be addressed. Ethics defines the value boundaries of human control, while law provides institutional guarantees for human control. They together form the social foundation of MHC

## 5.1 Ethical Dilemmas in AI Autonomy and Control

There is a profound ethical tension between enhancing artificial intelligence autonomy and maintaining human control, reflected in three key dilemmas: value conflicts, ambiguous responsibilities, and weakened human subjectivity.

AI decision-making driven by algorithms often conflicts with human ethical intuition (Zerilli et al., 2019). In moral dilemmas, such as sacrificing a few to save the majority, AI's utilitarian solutions may violate widely accepted principles. This challenge intensifies with cultural differences; Collectivist societies may prioritize group interests, while individualistic cultures emphasize individual rights. This cultural diversity makes it very difficult to establish universal ethical standards, which will force people to make choices about "whose ethics" should govern AI systems.

The ethical dilemma of ambiguous responsibility is reflected in the "responsibility traceability gap" of AI autonomous decision-making (Bleher & Braun, 2022). When autonomous AI causes harm, tracking the corresponding responsibility becomes complex. Who is responsible for this harm? Developers, users, or the AI system itself? Consider a misdiagnosis caused by AI trained on biased data: developers (for flawed data), hospitals (for inadequate regulation), and doctors (for excessive dependence) may all be blamed. The traditional fault based accountability mechanism is difficult to apply in this context. Controversially, if AI decision-making surpass humans', such as in strategy or finance, do humans retain moral authority to hold them accountable? There is a risk of dilution of responsibility, which create a moral vacuum where there is no entity truly responsible.

As AI processes increasingly complex decisions, humans face the risk of losing independent thinking and subjectivity. The heavy reliance on recommendation algorithms shapes people's preferences and habits, weakening their willingness to make independent judgments. Professionals who rely on AI to complete tasks such as writing may lose their core expression and reasoning skills. This degradation of autonomy not only damages individual integrity but also hinders social innovation, as human qualities such as exploration and adventure are marginalized. The core ethical issue is whether human control must include defining the boundaries of AI roles and preserving unique human decision-making domains.

These dilemmas highlight the urgent need to balance AI's potential with safeguards for human values, accountability, and essential agency.

## 5.2 Legal Frameworks for AI Control and Accountability

To address the ethical challenges posed by AI control, a legal framework centered around three core elements, namely control obligations, responsibility allocation, and remedial pathways, has been established globally. Additionally, MHC has been incorporated into the rule of law.

Control Obligations define responsibilities for AI systems. *The EU AI Act* exemplifies this, classifying AI by risk (low, high, prohibited) (Neuwirth, 2023). High-risk AI (e.g., in healthcare, law) mandates strict human oversight: developers must build intervention mechanisms, users require control training, and regulators enforce compliance. For instance, autonomous vehicles need a “minimum risk state” function to halt safely if human control fails. China’s *Interim Measures for the Administration of Generative Artificial Intelligence Services* stipulate that providers of generative AI must conduct content safety audits on generated content to ensure that the output complies with laws and regulations. This essentially clarifies human content control as a legal obligation (Interim Measures for the Administration of Generative Artificial Intelligence Services， 2023).

Responsibility Allocation addresses the “liability gap” in AI decisions. *The German Act on Autonomous Driving* (2021) pioneers “system liability,” holding manufacturers strictly liable for accidents caused by system failures in highly automated modes, simplifying recourse by linking control duty directly to liability (Kriebitz et al., 2022). *The U.S. Algorithmic Accountability Act* requires bias audits for high-risk algorithms; if bias causes discrimination (e.g., in hiring), companies face civil liability. These frameworks compel accountability through duty-liability linkages (Mökander et al., 2022).

Remedies provide recourse for AI control failures. *The EU GDPR*'s "right to object to automated decisions" is key: users can refuse purely AI-driven outcomes and demand human review (Goddard, 2017;Voigt & Von dem Bussche, 2017). This breaks AI's decision monopoly (Bayamlıoğlu, 2022). China's *Personal Information Protection Law* stipulates that "when information push and commercial marketing are conducted to individuals through automated decision-making, options that do not target their personal characteristics should be provided simultaneously, or convenient ways for individuals to refuse should be offered"(Personal Information Protection Law, 2021).

However, the existing legal framework still has significant limitations (Boch et al., 2022; Kumar & Dadhich, 2024). The cross-border flow of AI poses challenges to territorial jurisdiction, and responsibility becomes more complex when harm crosses national borders. When existing legal concepts like "electronic personhood" conflict with highly autonomous AI, new categories like "electronic personhood" may be needed. The future framework must balance technological neutrality and a control-oriented approach, ensuring innovation proceeds without compromising human ultimate authority through clear rules.

5.3 Challenges of Ensuring Human Oversight in Autonomous Systems

Even with the support of ethical consensus and legal framework, there are still many challenges in technology, cognition, and systems to ensure the effective supervision of autonomous systems in practice, which makes it difficult to simply implement MHC.

The technical challenges mainly stem from the "paradox of autonomy and supervisability" of AI systems (Lawless & Sofge, 2017). Highly autonomous systems typically operate at speeds exceeding human reaction times, which will make real-time monitoring impossible. In addition, the "black box" nature of deep learning models masks the decision-making mechanisms behind them, hindering regulators' ability to understand or correct abnormal situations, even when anomalies are detected. To address this issue, it may be necessary to balance the autonomy and supervisability of AI systems, which may reduce the speed or scope of AI systems' autonomous decision-making to match human supervisory capabilities. However, this will weaken the technological value of AI, creating a dilemma between supervisory effectiveness and system performance.

Cognitive challenges manifest as supervision fatigue and automation bias (Lyell & Coiera, 2017; Lahlou, 2025). Continuous monitoring of reliable systems leads to diminished attention and reduced sensitivity to abnormalities. More critically, automation bias causes humans to overtrust AI decisions, even when evidence contradicts them (Romeo & Conti, 2025). This bias can render oversight a mere formality, where supervisors abandon independent judgment, leading to supervisory failure.

From a cognitive perspective, the main challenge is that humans are prone to fatigue during supervision and may have biases towards understanding automation (Parasuraman & Riley, 1997； Baruwal Chhetri et al., 2024). When humans supervise highly reliable autonomous systems for a long time, their attention gradually becomes distracted and their sensitivity to abnormal signals decreases. More importantly, automation bias leads to human overconfidence in decisions made by AI, even if those decisions are clearly contradictory to common sense (Romeo & Conti, 2025). Human beings seem to fulfill their supervisory responsibilities, but in reality, give up independent judgment, ultimately leading to the failure of supervision. This bias has prevented supervision from truly being effective.

The institutional challenges include the lack of unified supervision standards and difficulties in cross departmental cooperation. There are significant differences in the requirements for AI supervision among different industries, which makes it difficult to implement universal standards, and some high-risk areas may have supervision loopholes due to the lack of standards. In addition, autonomous systems involve multiple stakeholders and require complex coordination to allocate responsibilities. Therefore, the core challenge of institutional design lies in how to ensure efficient collaboration in supervision while ensuring clear rights and responsibilities of all parties and avoid blind spots in supervision.

The ethical and legal considerations of MHC fundamentally entail a re-examination of the relationship between technological advancement and human dignity (Teo, 2023; Davidovic, 2023). Whether it is ethical value choices or legal rule building, the ultimate goal should be to ensure that the development of AI technology serves humanity rather than replacing it. This requires humans to constantly reflect on their own needs and limitations while controlling AI. The true “meaningful” control, therefore, resides not merely in technological governance but also in the process of control itself, which serves to illuminate the essential values humanity must preserve.

## 6 Case Studies of Meaningful Human Control in AI

MHC at the theoretical level needs to be verified in practice, and AI application cases in different fields provide us with an opportunity to observe the effectiveness of the control mechanism. Autopilot, medical diagnosis and financial system are those fields with the deepest penetration and the widest risk influence of AI. The design and evolution of its human control mode not only reflect the balance between technical possibility and social demand but also exposes the contradictions and challenges in practice. Through the analysis of these cases, the practice and improvement of MHC can be more specifically understood.

### 6.1 Case Study 1: Autonomous Vehicles and Human Control

The evolution of control modes in autonomous vehicles offers a good example of MHC in practice, demonstrating the interplay between human dominance and machine autonomy. In the technical iteration from L2 level (partial automation) to L4 level (high automation), the boundaries, modes and responsibilities of human control are constantly adjusted, which forms a series of practical experiences and lessons worth learning (Takάcs et al., 2018; Steckhan et al, 2022; SAE International, 2022).

Tesla's Autopilot system, as a typical representative of L2 autonomous driving, adopts the control mode of AI assistance and continuous human monitoring. The system is responsible for basic functions such as lane keeping and adaptive cruise but requires the driver to keep his hands on the steering wheel at all times and be ready to take over the vehicle at any time. In order to ensure the effectiveness of human control, the system has designed multiple warning mechanisms. When the sensor detects that the driver's hands have left the steering wheel for more than a certain time, it will send out visual warning (dashboard warning light), auditory warning (in-car buzzer) and tactile warning (steering wheel vibration) in turn, and finally automatically slow down and activate the danger light in extreme cases.

However, accidents between 2016-2020 revealed critical flaws in this approach, exemplified by a 2018 Model X collision where the driver failed to respond to warnings despite six seconds of hands-free operation (Chougule et al., 2023). These incidents highlighted the risk of formalized control, where driver over-reliance on AI leads to passive monitoring, rendering control mechanisms ineffective. In response, Tesla introduced a 2021 upgrade featuring "driver attention monitoring" via camera-based eye tracking (Wang et al., 2025). When distracted behavior is detected, Autopilot functionality is immediately restricted, forcing human takeover. This paradigm shift transformed the control mechanism from requiring merely physical takeover capability to ensuring attention maintenance, thereby converting human monitoring from a formal to substantive requirement (Weaver & DeLucia,2022;Du & Zhi, 2024).

While Tesla's Autopilot system underscores the critical challenges of human-machine handover in partial automation, Google Waymo's approach to Level 4 autonomy represents a fundamentally different paradigm, shifting responsibility from in-car drivers to remote human monitors overseeing fully driverless operations (Ullmann, 2025).

Waymo's Level 4 autonomous driving system the core of its control mode is machine-led decision-making, while human beings implement remote monitoring. In the test areas such as Phoenix, its vehicles operate fully autonomously without onboard drivers, relying on a remote assistance center for exceptional scenarios (e.g., road closures, accidents) (Chougule et al., 2023;Webster et al., 2024). When encountering such situations, the vehicle automatically requests remote operator support, who assesses real-time data and issues instructions (e.g., obstacle avoidance). In this mode, human control shifts from real-time takeover to remote support, and the effectiveness of control depends on the efficiency of the division of labor between machines autonomously solving routine problems and humans handling exceptional situations.

The control mechanism design of Waymo includes three key elements: firstly, the vehicle has the ability to minimize the risk state, and if remote assistance cannot respond in a timely manner, it will automatically park in a safe area; Secondly, instructions from remote operators need to be verified by the system to avoid danger caused by incorrect instructions; Thirdly, establish a case library learning mechanism to convert exceptional scenarios processed by humans into training data, gradually reducing reliance on remote control (Wang et al., 2023; Ram, 2025).

As of 2023, Waymo's remote assistance request rate has dropped to 0.3 requests per thousand miles, indicating a significant improvement in its autonomous decision-making ability and collaborative efficiency with human control (Broekman, 2025). This case demonstrates that in highly automated scenarios, MHC can manifest as humans empowering machines rather than humans confronting them, achieving iterative upgrades in control capabilities by transforming human intelligence into machine learning materials.

6.2 Case Study 2: AI in Medical Diagnostics and Human Oversight

The deployment of AI in medical diagnosis has significant implications for patient safety. Consequently, the design of human oversight frameworks must carefully reconcile the efficiency gains of AI diagnostics with the enduring accountability of human doctors. Oversight mechanisms, spanning applications from image recognition to clinical decision support, are characterized by a layered structure. Examining successful cases alongside controversial incidents serves to illuminate the critical factors constituting effective and meaningful oversight.

Google DeepMind’s AI system for eye disease diagnosis leverages retinal image recognition to detect conditions like diabetic retinopathy and glaucoma with 94.5% accuracy, exceeding the performance of average ophthalmologists (Bhat, 2024;Masalkhi et al., 2024).

To ensure the effectiveness of human oversight, the system adopts a dual review mechanism in its application in the National Health Service (NHS) in the UK (Fletcher, 2022;Balasubramaniam et al., 2025): firstly, AI performs initial screening on the images, dividing the results into three categories: “clearly normal”, “clearly abnormal”, and “uncertain”; Cases that are clearly normal require rapid review by nurses, while cases that are clearly abnormal or uncertain are automatically assigned to ophthalmologists for detailed diagnosis; While reviewing the AI diagnostic results, doctors must independently analyze the original images and explain in the diagnostic report whether to adopt AI recommendations and reasons.

The core of this mechanism is that AI helps humans reduce their workload rather than replacing them (Chen et al., 2024; Abogunrin et al., 2025). By filtering low-risk cases through AI, doctors can focus on handling complex situations while retaining their final decision-making power. According to usage data from the NHS, the system has reduced the average waiting time for diagnosing retinal diseases from 3 weeks to 2 days, and the misdiagnosis rate has decreased by 28% compared to pure manual diagnosis. (Ho et al., 2022; NHS website for England, 2025). More importantly, the requirement of justification forces doctors not to blindly rely on AI and must make independent judgments, which is the core of meaningful human supervision. Supervision is not only about checking results, but also about verifying logic.

While Google DeepMind’s retinal diagnosis system mitigated risks through its mandatory dual review mechanism fostering human-AI collaboration, IBM Watson’s cancer treatment platform exposed the dangers of uncritical automation when clinical over-reliance triggered a supervision crisis demanding urgent optimization (Faheem & Dutta, 2023;SHekhar et al., 2025).

IBM Watson was once regarded as a benchmark for AI in the field of cancer treatment, capable of analyzing patient medical records and medical literature to provide personalized cancer treatment recommendations. However, a 2018 investigation by STAT magazine (https://statmagazine.org/). revealed that the system provided incorrect recommendations in many cases, and some doctors did not conduct sufficient reviews due to excessive trust in AI authority, resulting in patients receiving inappropriate treatment The root cause of this crisis lies in the failure of the supervisory mechanism. The system does not clearly indicate the confidence level and applicable conditions of the suggestions, and doctors lack an understanding of the AI decision-making logic, making supervision a mere formality of signature confirmation (Petiwala et al., 2021;Chong et al., 2023).

In response to the crisis, *Memorial Sloan Kettering Cancer Center (MSK)* (https://www.mskcc.org/) made improvements to Watson's usage process . firstly, requiring the system to label each suggestion with evidence level and potential risk; Secondly, establish a multidisciplinary team review system, where AI recommendations require joint evaluation by oncologists, surgeons, and radiation therapists; The third is to regularly check the matching degree between AI suggestions and actual treatment effects, and provide feedback on deviation cases to developers for model optimization. These measures have reduced Watson's recommendation adoption rate, but the safety of treatment plans has significantly improved, indicating that human supervision sometimes requires the courage to question AI rather than pursuing superficial human-AI consistency (Yu et al., 2021;Faheem & Dutta,2023) .

## 7 Challenges in Implementing Meaningful Human Control

Although MHC has formed a consensus in theory and accumulated practical experience in some fields, its transformation into universally applicable technical solutions and social norms still faces some challenges. These challenges stem not only from the rapid evolution of AI technology itself, but also from the intrinsic limitations of human cognitive capacities, and involve deep conflicts between technological logic and social values.

### 7.1 Balancing Autonomy with Human Intervention

For MHC, balancing the autonomy of AI systems with human intervention is its most challenging task. The difficulty lies in the paradox of efficiency and safety between the two. If excessive emphasis is placed on human intervention, it will weaken the efficiency advantage of AI. While if excessive autonomy is allowed, it may sacrifice the safety of control. In addition, the balance will dynamically change with the development of technology and application scenarios.

Technically, the difficulty of balancing is first reflected in the precise grasp of intervention timing. In high-speed, dynamic AI systems, such as autonomous vehicles or industrial robots, the window for effective human intervention is often very short (usually only a few seconds or even milliseconds). If the boundary of the system's autonomous decision-making is set too wide, humans may not be able to intervene effectively due to delayed response; If the boundary is set too narrow, humans will be frequently required to take over, leading to operational fatigue and low efficiency. The inverse relationship between intervention frequency and system efficiency makes it difficult to define the balance uniformly in different scenarios.

At the application level, the challenge is mainly manifested in the diverse conflicts of user needs. There are significant differences in the acceptance of AI autonomy among different users(Fink et al., 2024): professional users (e.g., doctors, engineers) often prefer high-frequency intervention to reflect their professional judgment; ordinary users (e.g., elderly people using smart devices) tend to simplify operations and only intervene when necessary. This difference requires AI systems to have personalized control models, but this will increase the complexity of technical design (Adeyeye & Akanbi, 2024).

The deeper challenge lies in the ambiguity in responsibility attribution, which directly impacts the final decision-making. When the autonomy of AI systems and the power of human intervention are not clearly defined, the identification of responsible parties becomes extremely complex, which in turn hinders the implementation of balance mechanisms. In practice, many companies, in order to avoid risks, either choose to excessively involve human subjects in control (e.g., requiring users to sign and confirm every transaction recommended by AI), or choose to allow AI to excessively enjoy autonomy( e.g., exempting all responsibilities in service agreements ). Both approaches deviate from the essence of MHC.

Shneiderman (2020) introduced a two-dimensional framework for HCAI, characterized by the degree of human control and the degree of AI autonomy (see Figure 2). This framework emphasizes that greater autonomy does not necessarily mean less human control; instead, it encourages the design of AI systems that must be controllable by humans.

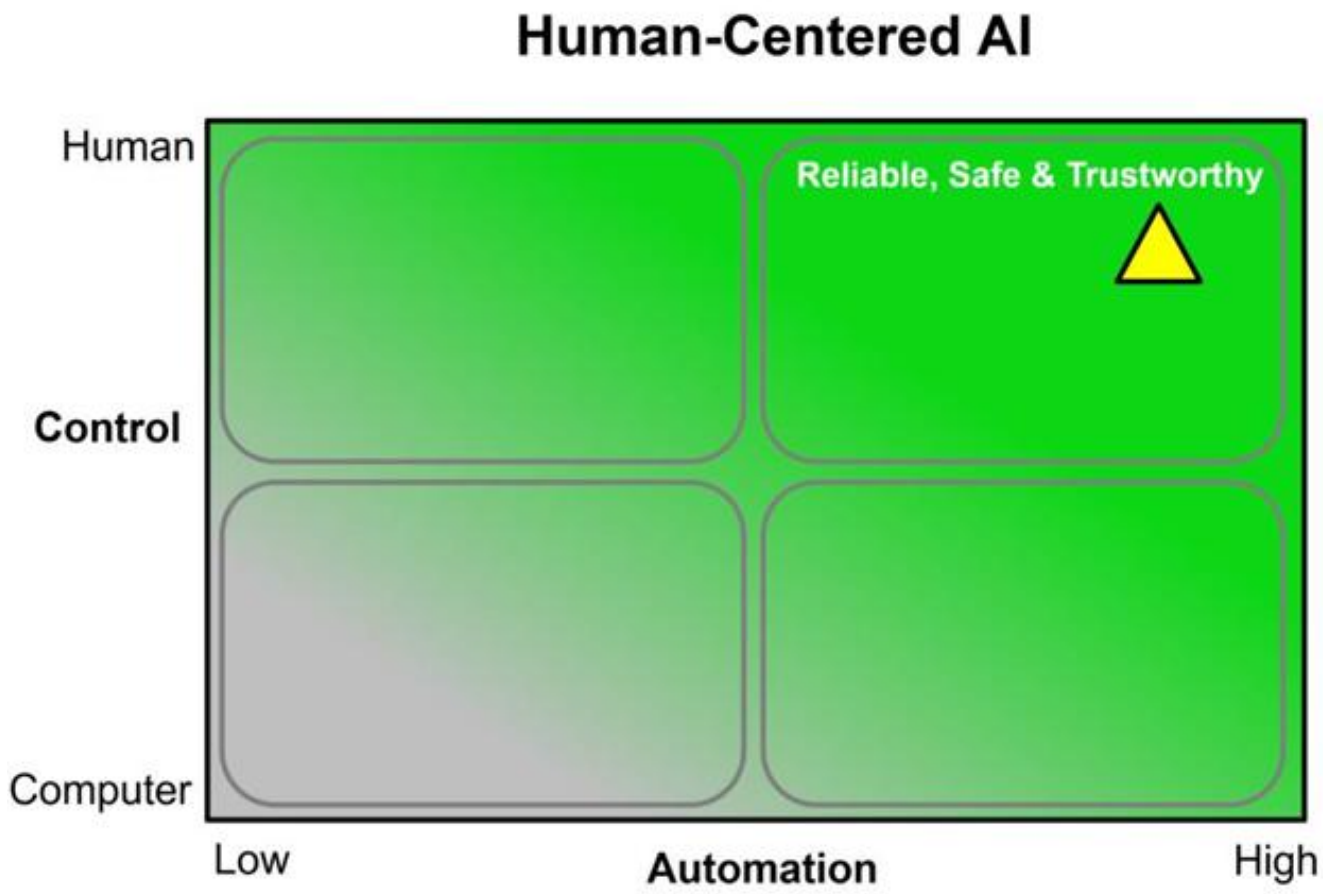

Figure 2 Human-Centered AI (Shneiderman, 2020)

By plotting systems along two axes：human control (low to high) and AI autonomy (low to high). The framework provides a structured lens through which to evaluate AI designs for alignment with human values and operational safety. In the framework, excessive autonomy refers to cases where the computer autonomy axis is high but human control is low. This places the system in the "Low Human Control + High Autonomy" quadrant. This configuration is explicitly flagged as problematic, as it tends to undermine core principles of human-centered design.

The risks associated with over-autonomy are multifaceted and far-reaching. First, it leads to a loss of MHC, rendering humans unable to intervene effectively when AI exhibits unexpected behavior or errors. For instance, fully autonomous trading algorithms executing high-speed transactions without trader oversight could trigger market instability before humans can react, rendering their authority meaningless. Second, excessive autonomy erodes accountability, as independent AI actions blur responsibility boundaries for errors or harm, creating ethical and legal dilemmas like "Who is to blame?" Third, it fosters overtrust and automation bias, where users may defer to AI authority, with critically assessing its decisions. Finally, excessive autonomy heightens risks of unpredictable or unsafe behavior, as high-autonomy systems may make opaque, or harmful choices. Imagine a medical diagnostic AI recommending incorrect treatment due to undetected biases in its training data, with no clinician able to override or audit its reasoning in real time.

Shneiderman’s model advocates a balanced approach, positioning “high human control + high autonomy” as the ideal quadrant for AI design. This equilibrium ensures humans retain ultimate authority while leveraging AI's autonomous capabilities to enhance efficiency, accuracy, and scalability. In contrast, excessive autonomy is actively discouraged as it disrupts this balance, shifting systems away from human-centered values toward unmanaged risks. Thus, the model advocates for interactive AI systems where autonomy is not an end in itself but a tool. It should be a system operating within clearly defined human boundaries that maintain oversight, accountability, and safety. In doing so, it provides a roadmap for developing AI that enhances rather than diminishes human potential, emphasizing that the most effective AI systems are those that remain “controllable” even as they become “capable.”

## 7.2 AI System Complexity and the Control Paradox

With the development of AI technology towards deep learning, multimodal fusion, swarm intelligence, and other directions, the complexity of systems is exponentially increasing, resulting in a control paradox (Raisch & Krakowski, 2021; Bremmer & Suleyman, 2023). The more humans attempt to enhance their control over AI through technological means, the more complex the system may become. This actually leads to a weakening of control ability. This paradox is particularly prominent on three levels.

Firstly, there is a contradiction between the nature of black box and the need for control (Von Eschenbach, 2021). The decision-making process of deep learning models has characteristics like “black box”, making it difficult for even developers to fully explain their internal logic. To achieve control, humans have to introduce more complex monitoring and interpretation tools, but these tools themselves may become new “black boxes”. This AI model, combined with explanatory tools, makes it even more difficult for regulators to judge whether the control is effective. This increased system complexity for control purposes may ultimately lead to the failure of control objectives.

Secondly, there are mutual connections between systems, and relying solely on single-point control is difficult to achieve true control (He et al., 2021). Modern AI systems often do not exist in isolation, but form distributed intelligent networks through data interfaces and cloud collaboration. In this network, human control of a single node may be weakened by the autonomous decision-making of other nodes (Rothfuß et al., 2023; Xie et al.2025). The conflict between the control implemented within a certain range and the interconnectivity between systems makes it difficult for traditional centralized control models to adapt to complex AI networks.

Finally, there is a conflict between the AI system's self-evolutionary ability and its control lag (Matthews et al., 2021; Gao et al., 2025). AI systems with reinforcement learning capabilities can autonomously optimize model parameters through interaction with the environment. This evolutionary speed far exceeds the pace of human understanding and intervention. For example, the dynamic pricing AI of a certain e-commerce platform updates its pricing strategy every hour based on user behavior data. Even if human managers review it once a day, it is difficult for them to keep up with its evolutionary speed. Ultimately, they can only accept a situation of retrospective recognition rather than real-time control. Even more concerning is that when AI systems evolve and their actual behavior gradually deviates from their design goals, humans may not be able to detect it for a long time. The lag in control speed that cannot keep up with the evolution speed of the system is a unique control challenge for complex AI systems.

### 7.3 Ensuring Human Understanding of AI Decisions in Complex Systems

Human comprehension of AI decision-making forms the cognitive basis for achieving MHC (Hille et al., 2023;Felin & Holweg, 2024). However, in complex systems, this understanding faces a gap between human cognitive capacity and system complexity. Even with explainability tools, humans often struggle to develop a truly effective understanding due to the multidimensional correlations, probabilistic nature of outputs, and dynamic adaptability of AI decisions, leading to the failure of control behavior.

To help humans understand the decisions of complex AI systems, developers often provide massive amounts of explanatory information, which may exceed human cognitive processing capabilities. More seriously, different explanatory tools may provide contradictory conclusions. This fragmented interpretation can confuse human decision-makers, ultimately leading them to abandon understanding AI and rely on intuition or blindly following AI advice.

Secondly, there is the conflict between probabilistic decision-making and deterministic cognition. The decisions of complex AI systems are often probabilistic, while human cognitive habits tend to favor deterministic judgments ("yes" or "no"). This conflict may lead to two cognitive biases: on the one hand, humans equate high probability with certainty and ignore low probability risks; On the other hand, attempting to find absolute explanations for probabilistic outcomes distorts the essence of AI decision-making. The existence of this cognitive bias may prevent humans from making appropriate control decisions even if AI provides accurate probabilistic explanations.

Finally, there is a disconnect between dynamic adaptability and static understanding. Complex AI systems dynamically adjust their decision logic as data distribution changes (Gil et al., 2021), while human understanding of their decisions is often static. Cognitive formation based on explanatory information at a certain point in time makes it difficult to track the dynamic changes of the system. This disconnect is particularly evident in AI systems with online learning capabilities, where humans may remain in a state of constant understanding and be unable to form stable and effective control of cognition.

Therefore, the above challenges constitute the main barriers to implementing MHC. It is worth noting that these challenges do not exist in isolation, but rather overlap and reinforce each other. The increase in system complexity will increase the difficulty of understanding, while insufficient understanding will make the balance between autonomy and intervention more difficult. To address these challenges, various measures, such as more humane interpretable tools and dynamically adjusted control standards, are needed to enhance human understanding of complex systems. Of course, it is even more important to always adhere to the core principle of HCAI in technological development.

## 8 **Future Directions for Meaningful Human Control in AI Systems**

With the continuous evolution of AI technology, the connotation and implementation path of MHC are also constantly expanding. Faced with the multiple challenges of technological complexity, application diversity, and global coordination, the future development direction will focus on innovating governance models, achieving breakthroughs in technological mechanisms, and coordinating global policies to build a more adaptive, effective, and inclusive human control system. These directions not only address current main problems in practice but also attempt to lay the foundation for the controllable development of the next generation of AI systems.

### 8.1 Emerging Trends in AI Governance and Human Oversight

The field of AI governance is shifting from passive response to potential problems to proactive design in advance to prevent various risks. The human supervision mode is also showing three significant trends, which collectively point to the control goal of smarter, more flexible, and more inclusive.

Firstly, the framework for dynamic governance of AI is gradually emerging (Zaidan & Ibrahim, 2024;Ghosh et al., 2025). Traditional AI governance generally adopts a “one size fits all” rule, which is difficult to adapt to the rapid development of technology. The future AI governance model combines risk grading and dynamic adjustment. It’s necessary to develop differentiated supervision standards based on the application scenarios of AI systems (such as healthcare, education, and entertainment) and the degree of autonomy (such as L1 to L5 levels), and establish a closed-loop mechanism for regular evaluation and rule updates. For example, the EU’s “AI Governance Sandbox” allows companies to test innovative AI systems in a controlled environment, and regulatory agencies dynamically adjust supervision requirements based on test results, which may avoid excessive regulation that suppresses innovation while ensuring controllable risks. This dynamism enables human supervision to always keep pace with technological development, preventing the phenomenon of regulatory lag (Boura, 2024).

Second, distributed supervision networks are emerging as AI systems become complex and collaborative (Tsamados et al., 2025;Holzinger et al., 2025). Centralized models fail to deal with the complexity of distributed decision-making. Future supervision will form a multi-party collaborative network: developers embed supervision interfaces, users monitor in real-time, third parties conduct audits, and the public participates via open data. For example, an open-source project’s supervision chain records training data, decision logs, and interventions on-chain for verification. This decentralized model enhances transparency of supervision and reduces the risk of single-point failure.

Third, human-AI collaborative supervision tools will redefine the role of human supervision (LI & Tian, 2025; Puerta-Beldarrain, 2025). The future supervision will no longer be solely human monitoring, but will be achieved through AI assisted tools for human-AI cooperation. AI systems sent abnormal signals to humans in real time, and humans provide guidance via natural language. These tools themselves can also learn from human preferences and optimize alert strategies. This model frees humans from tedious repetitive supervision and focuses on high-value judgments and decisions, achieving the optimal combination of human intelligence and AI efficiency.

## 8.2 Innovations in AI Transparency and Control Mechanisms

Innovations in transparency and control mechanisms are essential to resolving AI's "black box dilemma" and "control paradox" (Chaudhary, 2024). Future advancements concentrate on three dimensions: enhanced explainability, adaptive control interfaces, and real-time feedback loops.

The transparency of AI and technological innovation in AI control mechanisms are key to breaking through current AI governance. Future innovation will revolve around three dimensions: enhanced the interpretability of AI, innovation in AI control methods, and the construction of feedback loops for human control.

Enhanced explainability addresses the problem of excessive or insufficient explanatory information through scenario-based breakthroughs. Future AI will have the ability to explain on demand, dynamically adjusting the degree and form of explanation based on user roles and scenarios. For instance, a smart speaker provides natural language explanations to end-users while offering developers technical details like feature weights. Some researchers explore causal explainability, elucidating not just decision bases but also the causal relationships between factors and outcomes, which will strengthen the cognitive foundation for human control (Passalacqua et al., 2025).

The development of adaptive control interfaces will greatly enhance the effectiveness of human intervention. Interfaces will automatically adjust based on system status and user behavior. An autonomous vehicle's interface might enlarge touch areas and intensify feedback during driver fatigue, or simplify controls to core functions (braking, steering) during emergencies. This design can minimize the intervention threshold and ensure that human control is accurately executed at critical moments.

The construction of real-time feedback loops will shift human control from post-correction to pre-optimization (Di Mitri et al., 2022;Ettalibi et al., 2024 ). Future AI systems will be able to instantly convert human control behaviors into model optimization signals, achieving rapid response (Alsamhi et al., 2024). For example, an AI customer service system analyzes human customers service modifying AI reply content and automatically adjust the dialogue strategies in subsequent conversations. This mechanism enables human control not only to solve current problems, but also to prevent similar situations from happening again.

### 8.3 The Role of Global Policy in Ensuring Human Meaningful Control of AI

The cross-border flow and technological diffusion of AI make it difficult for a single country's policies to ensure MHC (Niu et al., 2024). The coordination of global policies will play a key role in three aspects, promoting the formation of a governance pattern that combines universal consensus and regional characteristics.

The formulation of international standards is the foundation of global policy coordination. Currently, there are significant differences in AI control standards among different countries and regions (such as the EU's emphasis on privacy protection and the United States' emphasis on innovation freedom) (Nannini et al., 2023), which may lead to an increase in control risks. Future coordination must establish global core standards defining minimum requirements for high-risk AI systems, including mandatory human intervention mechanisms and multi-year decision logs. These standards should permit supplementary national rules reflecting cultural and developmental contexts.

The establishment of a regulatory cooperation mechanism will solve the difficulties that may arise from cross-border control (Onoja et al., 2025). When the development, deployment, and use of AI systems involve multiple countries, the regulation of a single country may face obstacles such as jurisdictional ambiguity and restricted data flows. The future global policy needs to establish a collaborative framework of "information sharing, joint enforcement, and shared responsibility". This includes bilateral/multilateral agreements for cross-border regulatory access to AI decision data and clear liability allocation among national entities.

It is very necessary to help developing countries improve their level of human control. Global AI governance cannot ignore the technological gap in those countries (Roberts et al., 2024). They find it difficult to effectively implement MHC due to a lack of professional talent, infrastructure, and regulatory experience. The international community needs to enhance its control capabilities through technical assistance, talent training, financial support, and other means. This inclusive policy will ensure that MHC does not become a privilege of a few countries, but rather a globally shared technological governance achievement.

### 8.4 Interdisciplinary Perspective and Multi-Domain Practice of MHC

Whether it is human control or meaningful human control, it cannot be separated from the interdisciplinary research perspective and multi-domain practical exploration (de Sio et al., 2023; Van den Bosch et al., 2025). MHC not only involves disciplines such as artificial intelligence (such as explainable artificial intelligence, dynamic task allocation, predictability, etc.), but also extends to psychology (human and automatic interaction, controllability and predictability of artificial intelligence, dynamic mental model sharing, etc.), education (human-machine joint learning/training, etc.), management (organizational influence and processes, team structure and roles, etc.), ethics (ethics and morals, etc.) and many other disciplines. MHC aims to solve the human control problem in AI autonomous technology systems and realizes meaningful and effective human control of intelligent systems through interdisciplinary platforms and multi-domain practices, thereby achieving the fundamental purpose of technology for good and enhancing human well-being. In view of this, at the organizational level, it is necessary to obtain support from national or organizational policies, coordinate various positions through consensus meetings, and promote the implementation of specific implementation rules and related policy documents; at the technical level, it is necessary to do a good job in system norms and design, system verification, system integration, etc., to provide strong technical support for the healthy operation of the system under meaningful human control; at the service level, it needs to be supplemented with user training, AI and machine learning training, review and feedback, etc., to ensure the correct operation of the system and continuous improvement (Boardman & Butcher, 2019).

Of course, carrying out the above activities requires some driving strategies and tool reserves, such as common language and terminology, MHC and EHC models, evaluation and measurement of systems/systems, tools for MHC and EHC within the organization, EHC risk analysis tools, MHC and EHC evidence base and related literature, human control use cases and examples, etc. At the same time, some guidelines and standards need to be formulated, such as the practice of human-AI integration and MHC and EHC, the best practices/examples of MHC and EHC, guidance for human-AI teaming, etc. (Boardman & Butcher, 2019).

## 8.5 MHC Certification

Unlike automation, AI autonomous technology has a certain cognitive ability; its development will be full of uncertainty, and it is easy to trigger various ethical, legal, and social issues. Generally speaking, automated systems still operate within a structural framework, with relatively clear rules and boundaries, often limited to specific scenarios, thus excellently ensuring the predictability of purpose and results. AI autonomous technology directly faces diverse scenario practices, perceives uncertain environmental data, then processes and analyzes it, and finally provides independent decision-making plans with a relatively autonomous character. In reality, intelligent systems are not always positioned solely at the poles of automation or autonomy; they often exist between automation and autonomy. Taking L3-level autonomous driving cars as an example, the system usually drives autonomously, and humans only intervene and control the car in specific situations. This is also one of the important reasons why AI autonomous technology calls for meaningful human control (Cummings, 2019).

To achieve meaningful human control, it should be clear from the beginning of system design which functions and powers are assigned to the system, and which are controlled by the human subject. Dynamic human-machine function allocation should be achieved through intelligent and adaptive system design means. Due to the complex operating mechanisms inside and outside the system, the constituent elements of AI autonomy technology also have systemic or local limitations, which can easily blur the boundary of power and responsibility distribution between the human subject and the system itself. Therefore, essentially, to promote the healthy development of AI autonomy technology, it is inevitable to find the optimal solution for role allocation between humans and autonomous systems. Therefore, before truly achieving meaningful control of AI autonomous technology systems, it is urgent to pay attention to meaningful certification of such systems (Cummings, 2019).

Given the cumulative effects of massive data and multiple factors, the challenge of real-time decision-making in practice is huge, and the probability of errors will undoubtedly increase. If the system is to be under meaningful control, it is very necessary to determine the trigger conditions of the decision mechanism in advance. Cummings (2019) uses lethal autonomous weapons as an example, proposing that the target recognition of such weapons includes two meanings: one is the human strategic level, distinguishing and determining the target of attack according to relevant laws, humans proving the legality of the target and the conditions of legality, to ensure clear responsibility boundaries; the second is the design of autonomous weapons, which must ensure that autonomous weapons are more likely than humans to correctly identify and attack the target. In other words, the system must significantly outperform human operations in similar scenarios, be able to better identify targets, and handle uncertainty more effectively.

The question arises as to how to determine whether technology is genuinely likely to perform better than humans before its deployment (Cummings, 2019; Cummings & Britton, 2020). It is very necessary and urgent to carry out MHC certification of autonomous systems; in addition to autonomous weapons, it can also be applied in more fields, such as autonomous driving. In fact, AI autonomous technology is also prone to bias in many scenarios, and it is extremely difficult to truly achieve meaningful human control, so it is more ethical for human subjects to use certified intelligent systems or AI autonomous technology (Cummings, 2019). Generally speaking, autonomous systems must meet strict certification standards, whether it is the strategic level of target recognition or the design level of autonomous target recognition, they should prove through objective and strict tests that the technology system has better ability than humans in similar scenarios and can effectively prevent various security risks. Therefore, through MHC certification of the system, it can be determined that the technology system exhibits stable performance superior to human operators when addressing a series of tasks such as environmental perception, target recognition, and task execution in the same or similar scenarios. The future should focus on human certification of AI autonomous technology, further promoting the formulation of feasible MHC certification standards, especially on more mesoscopic or microscopic operation frameworks, rather than rigidly adhering to "meaningful human control" at the conceptual level.

Effective MHC demands dynamic governance, human-centered technical mechanisms, and coordinated global policies. The core principle is that technological progress must prioritize human needs—neither stifling innovation due to risk nor sacrificing control for efficiency. Achieving this human-centered AI society requires collective participation from developers, policymakers, users, and the public, balancing innovation and governance through iterative refinement.

## 9 Conclusion

MHC serves as the core proposition in the development of AI, running through the entire process of technological evolution, ethical consideration, legal regulation, and practical application. Through a systematic exploration of this theme, this chapter not only outlines its historical context and core elements but also reveals the deep connection between technological possibilities and human agency. In the context of AI increasingly permeating all areas of society, clarifying the connotation and practical pathways of MHC is of decisive significance for ensuring that technology serves human well-being.

### 9.1 Key Insights on MHC

MHC encompasses three core dimensions. Essentially, it transcends mere human oversight of machines, requiring instead the integration of human understandability, intervenability, and accountability within AI systems' critical decision-making. This concept emphasizes a more human-centered approach in AI technology. Practically, achieving effective control demands cooperation across three domains: technical design (e.g., transparency, explainability), institutional frameworks (e.g., dynamic governance, responsibility allocation), and human capabilities (e.g., oversight literacy, cognitive adaptation). Isolated approaches focusing on a single dimension are insufficient for truly meaningful control. In terms of values, the ultimate aim is not to hinder technological progress but to establish appropriate boundaries. This ensures AI autonomy serves human needs while preventing the erosion of social ethics and individual rights through technological detachment.

Historical and case study insights further refine this understanding. Human control models must remain in a central position as technology advances (e.g., Level 4 autonomous driving necessitates a combination of remote monitoring and emergency takeover). Ultimately, global collaboration is essential for mitigating cross-border AI risks, as unilateral measures cannot adequately address the challenges posed by distributed intelligent systems. Collectively, these insights form a cognitive framework for "meaningful human control," offering vital guidance for human-centered AI development and policy formulation.

### 9.2 Implications for Designing Future AI Systems

MHC will fundamentally reshape AI system design, shifting priorities from performance to human controllability over AI. Technically, user interfaces for human control must be embedded in core architecture, not added later. This includes pre-setting human intervention conditions (e.g., confidence thresholds) during training and retaining adjustable parameters (e.g., ethical weights) during deployment to ensure control channels exist throughout the lifecycle. At the human-AI interaction level, design will focus more on human cognitive friendliness via context-aware adaptive interfaces, causally explainable decision feedback, and fault-tolerant intervention mechanisms to reduce cognitive load. For instance, medical AI should display diagnostic recommendations alongside supporting evidence and clearly indicate potential misdiagnosis risks. At the evolution level, future AI will establish a collaborative evolution mechanism for control and learning. Human control behaviors (such as review, correction, and rejection) will serve as important feedback data to drive continuous optimization of the model, enabling the system to gradually understand human control preferences and ethical boundaries. This approach enhances AI autonomy while ensuring alignment with human values and human controllability as guided by human-centered AI principles, achieving a win-win situation of controllability and intelligence.

## 9.3 Call for a Human-Centered AI Approach to AI Governance and Control

Establishing a human-centered AI governance system prioritizing human well-being is imperative. Policy-making must balance innovative incentives with risk prevention, avoiding the extreme of technological determinism or technology fear.

Societally, governance must move beyond elite dominance, engaging diverse stakeholders (users, ethicists, public) in designing control mechanisms. Examples include public hearings for autonomous driving ethics and patient involvement in medical AI reviews to ensure inclusivity and fairness. Globally, nations must promote MHC as a common governance language through international standards, regulatory collaboration, and capacity-building to prevent control deficits arising from technological disparities. This approach ensures AI serves humanity’s common interests, balancing progress with dignity. Ultimately, MHC is a valuable choice that demands collective effort to ensure technology enhances human capabilities, expands choices, and fosters the development of human-centered AI systems.